\documentclass[aps,prb,preprint]{revtex4}
\usepackage{amsmath}
\usepackage{graphicx}   

\def\be{\english\begin{equation}}

\def\ee{\end{equation}\farsi}

\def\bea{\english\begin{eqnarray}}

\def\eea{\end{eqnarray}\farsi}

\def\ds{\displaystyle}

\def\be{\begin{equation}}
\def\ee{\end{equation}}
\def\bea{\begin{eqnarray}}
\def\eea{\end{eqnarray}}
\begin{document}

\title{The point of departure  of a particle  sliding on a curved surface}

\author{Amir Aghamohammadi}
 \affiliation{Department of Physics, Alzahra University, Tehran 19384,
IRAN}

 \email{mohamadi@alzahra.ac.ir}
\date{\today}

\begin{abstract}
A particle is thrown tangentially on a surface. It is shown that for some surfaces and for special initial velocities
the thrown particle leaves immediately the surface, and for special conditions it never leaves the surface. The conditions for leaving the surface is investigated. The problem is studied for a surface with the cross section $y=f(x)$. The surfaces with the equations $f(x)= -\alpha x^{k}\
(\alpha, k>0)$ is considered in more detail.  At the end the effect of friction is also considered.
\end{abstract}

\maketitle

\section{Introduction}
One of the standard problems in elementary mechanics is finding the point where an sliding particle on a frictionless sphere lose contact with the sphere \cite{Hall,KLep,Irodov,Morin}. A particle at the top of a sphere with the radius $R$ is thrown tangentially with the velocity $v_0$. The particle leaves the sphere immediately provided that  $v_0\geq v_c:=\sqrt{Rg}$. For velocities less than $v_c$ it will move on the sphere and leave it later. A more realistic  variation to this problem is to consider what happens when kinetic friction is present. This problem have been solved exactly\cite{Mungan}. The problem of sliding  a particle on a  sphere with friction by a perturbation expansion in the coefficient of sliding friction and also by an exact integration of the equation of motion is studied\cite{PriorMele}.

The motion of a particle on an arbitrary surface, provided that the particle's path  is always in a vertical plane, is addressed in this article. It is shown that for some  surfaces there is no initial velocity for which the particle leaves the surface immediately, and there are  some other surfaces for which the thrown particle with any arbitrary nonzero initial velocity immediately leaves the surface. There are a surface  for which depending on initial velocity the particle leaves the surface immediately or it never leaves the surface. And finally there are surfaces for which the particle never leaves the surface for any arbitrary initial velocity.
The problem is studied for a surface with the cross section $y=f(x)$. The case $f(x)= -\alpha x^{k}\ (\alpha, k>0)$ is considered in more detail.  At the end the effect of friction is also considered.
The problem addressed in this article may be of interest in the framework of an undergraduate course in mechanics. 

\section{particle sliding on a parabolic surface}
As a simple example, let's first consider the motion of a particle on a frictionless parabolic surface with the cross section $y=-\alpha x^2$, where $y$ axis is vertical and the $x$ axis is horizontal. See figure 1. The particle's initial velocity is taken to be $ v_0>0$, and it is thrown  in the $x$ direction from the origin. Let's consider a circle whose radius is $R$ passing through the origin, and is tangent to the parabola.
The equation of the circle is
\begin{equation}
x^2+(y+R)^2=R^2.
\end{equation}
At the vicinity of the origin the equation of the circle is
\begin{equation}
y= -R+\sqrt{R^2-x^2}\approx -\frac{x^2}{2R}.
\end{equation}
If the second derivative of the two curves at the origin are the same, then the radius of the curvature of the parabola is the same as the radius of circle($R$). So the radius of the curvature of the parabola at the origin is $R= \ds{\frac{1}{2\alpha}}$. The Newton equation of motion for the particle at the origin
is
\begin{equation}
mg-N=\ds{\frac{mv_0^2}{R}}=2\alpha mv_0^2,\quad \Rightarrow \quad N=m(g-2\alpha v_0^2),
\end{equation}
where $N$ is the normal force exerted by the surface.  $N$ is positive provided that  $v_0$ be less than the critical velocity  $v_c:= \ds{\sqrt{\frac{g}{2\alpha}}}$.
What happens if $v_0=v_c$?
Let's assume there is no surface. Then the particle's path is a parabola with the equation
$y=-\ds{\frac{gx^2}{2v_0^2}}=-\alpha x^2$. So, if $v_0=v_c$, the particle moves tangent to the
parabola and $N$ is always equal to zero.
If the initial velocity $v_0>v_c$, the particle immediately
leaves  the surface, and if $v_0<v_c$ the particle never leaves the surface.
\begin{figure}
\begin{center}
\includegraphics{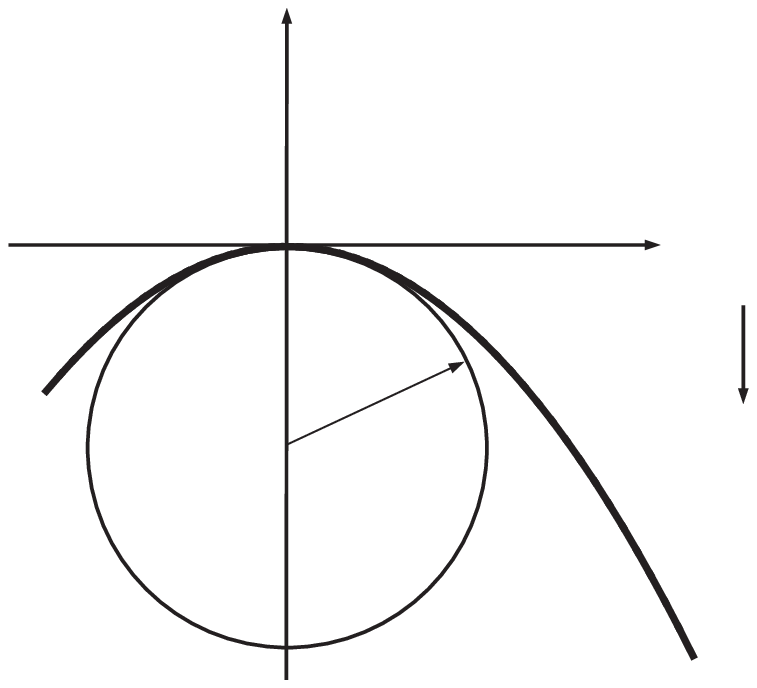}
\setlength{\unitlength}{1mm}
\put(-50.5,73.5){$y$} \put(-2,33.7){$g$}\put(-11,47){$x$}\put(-43,33){$R$}
\caption{\label{fig:inclinedplane}}{A particle sliding on a parabolic plane }
\end{center}
\end{figure}
\section{particle sliding on a surface whose cross section is $y=f(x)$}
Let's consider the motion of a particle on a frictionless surface whose cross section is $y=f(x)$, where $f(0)=0$. The particle is thrown  from the origin with the velocity $v_0$ tangential to the surface. Let's assume the particle leaves the surface at the point ${\bf r_d}$. At the point of the departure the force exerted by the surface should be equal to zero. Then at the point ${\bf r_d}$
the components of the acceleration are
\begin{equation}\label{05}
\ddot x\vert_{\bf r_d}=0,\qquad \ddot y\vert_{\bf r_d}=-g.
\end{equation}
Using the equation of the surface cross section $y=f(x)$, as long as the particle moves on the surface,
one may obtain
\begin{align}\label{06}
  &\dot y=\dot x f'(x)\nonumber \\
  &\ddot y=\ddot x f'(x)+\dot x^2 f''(x),
\end{align}
where prime means differentiation with respect to $x$. Using (\ref{05}) and (\ref{06}) at the point of departure one arrives at
\begin{align}\label{07}
  &\dot x\vert_{x_d}=\sqrt{\frac{g}{-f''(x_d)}} \\
  &\dot y\vert_{x_d}=f'(x_d)\sqrt{\frac{g}{-f''(x_d)}}.
\end{align}
As it may be expected at the point where the particle lose contact, $f''(x_d)$ should be negative. So the surfaces for which at all points $f''(x)>0$, there is no point of departure.
If the  particle's velocity at the point $x_d$  is $v_d$ then it will leave the surface
\begin{align}\label{07-2}
  v_d=\sqrt{\frac{g(1+f'^2(x_d))}{-f''(x_d)}}.
\end{align}
If the particle is thrown  from the origin with the velocity $v_0$ tangential to the surface, and provided that the surface is frictionless, conservation of energy gives
\begin{align}\label{08}
v^2=v_0^2-2g f(x).
\end{align}
Using (\ref{07}-\ref{08}), knowing the function $f(x)$, one may obtain the point of departure ${\bf r_d}$ from
\begin{equation}\label{09}
f''(x_d)(v_0^2-2gf(x_d))+g(1+f'^2(x_d))=0.
\end{equation}
The smallest solution  of this equation for $x_d$, if there exists any, gives the point of departure.
The function that solve (\ref{09}) for all $x_d$ is a parabola, subject to the projectile motion. It should be expected since (\ref{09}) is obtained setting $N=0$.

As an example let's consider the standard problem of finding the point where an sliding particle on a frictionless sphere lose contact with the sphere. Then
\begin{equation}\label{09-2}
x^2+(y+R)^2=R^2,\quad \Rightarrow\quad y=-R+\sqrt{R^2-x^2},
\end{equation}
where $R$ is the radius of the sphere and origin is taken to be at the top of the sphere. Using (\ref{09}), one arrives at
\begin{equation}\label{09-3}
y_d=-\frac{R}{3}+ \frac{v_0^2}{3g}.
\end{equation}

\subsection{example; $f(x)=-\alpha x^k,\ (\alpha, k>0) $}
Let's take $f(x)=-\alpha x^k,\ (\alpha, k>0)$.
We want to obtain the point of departure for different values of $k$.
See figure 2. Equation (\ref{09}) recasts to
\begin{equation}\label{10}
\alpha^2 g k(k-2)x_d^{2k-2}+\alpha v_0^2k(k-1)x_d^{k-2}=g.
\end{equation}
$x_d$ is point where the curve $A(x):= \alpha^2 g k(k-2)x^{2k-2}+\alpha v_0^2k(k-1)x^{k-2}$ crosses the line $B(x):=g$.
\begin{figure}
\begin{center}
\includegraphics{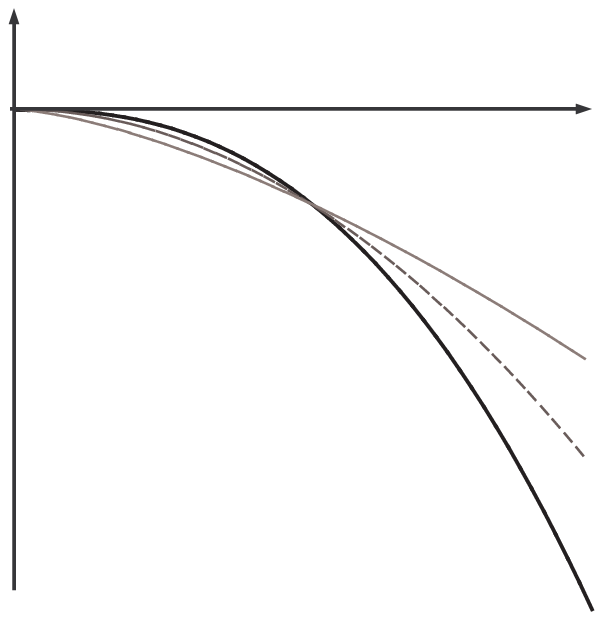}
\setlength{\unitlength}{1mm}
\put(-63,66){$y$} \put(-1.6,53.5){$x$}\put(-0.6,28.7){$1<k<2$}\put(-0.6,17){$k=2$}\put(-0.6,3){$k>2$}
\caption{\label{fig:inclinedplane}}{The surfaces with the cross section $y=-\alpha x^k$. }
\end{center}
\end{figure}
\begin{figure}
\begin{center}
\includegraphics{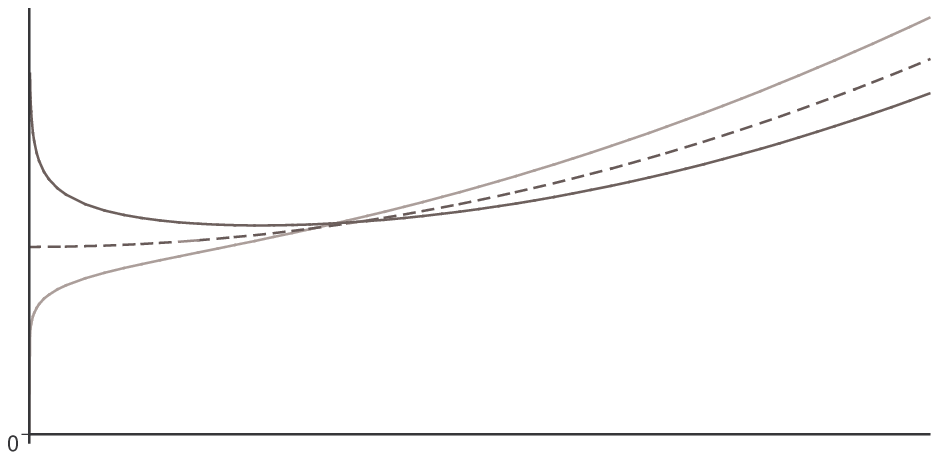}
\setlength{\unitlength}{1mm}
\put(-96,52){$v_d$} \put(-1,4.5){$x_d$}\put(-1.6,46.7){$1<k<2$}\put(-1.6,43){$k=2$}\put(-1.6,39){$k>2$}
\caption{\label{fig:inclinedplane}}{minimum velocity needed to leave the surface, $v_d$, versus $x_d$. }
\end{center}
\end{figure}
\subsubsection{$k>2$}
For the case $k>2$, $f''(0)=0$, and the radius of curvature at $x=0$, is infinite. Then
at the origin for any arbitrary tangential initial velocity, $N=mg + m\ddot y>0$. So it is not possible to throw the particle with any bounded initial velocity such that it leaves immediately the surface. If $v_0=0$, the particle remains at rest at the origin. However $x=0$ is an unstable point.

For $k>2$, $A(0)=0$, and $A(x)$ is an increasing function of $x$. So the curve $A(x)$ should cross
the line $B(x)=g$.  For any arbitrary small value of initial velocity $v_0\ne 0$, the particle will leave
the surface later. In figure 3 using (\ref{07-2}) the minimum velocity needed to leave the surface, $v_d$, is plotted in terms of $x_d$. As it is seen at the origin $v_d$ approaches to infinity which supports the above arguments. It is interesting to note that for $k>2$, $v_d$ has a minimum in terms of $x_d$. It can be easily shown that at the point on the surface whose $x$ component is $\left(\frac{k-2}{\alpha^2 k^3} \right)^{1/(2k-2)}$, $v_d$ is minimum.

There is another simple way to analyze the problem.  The intersection of two plots for $v_d$ and $v=\sqrt{v_0^2-2gf(x)}$ gives the point of the departure. In figure 4 three different initial velocities are considered. Particles with smaller initial velocities leaves the surface later.
\begin{figure}
\begin{center}
\includegraphics{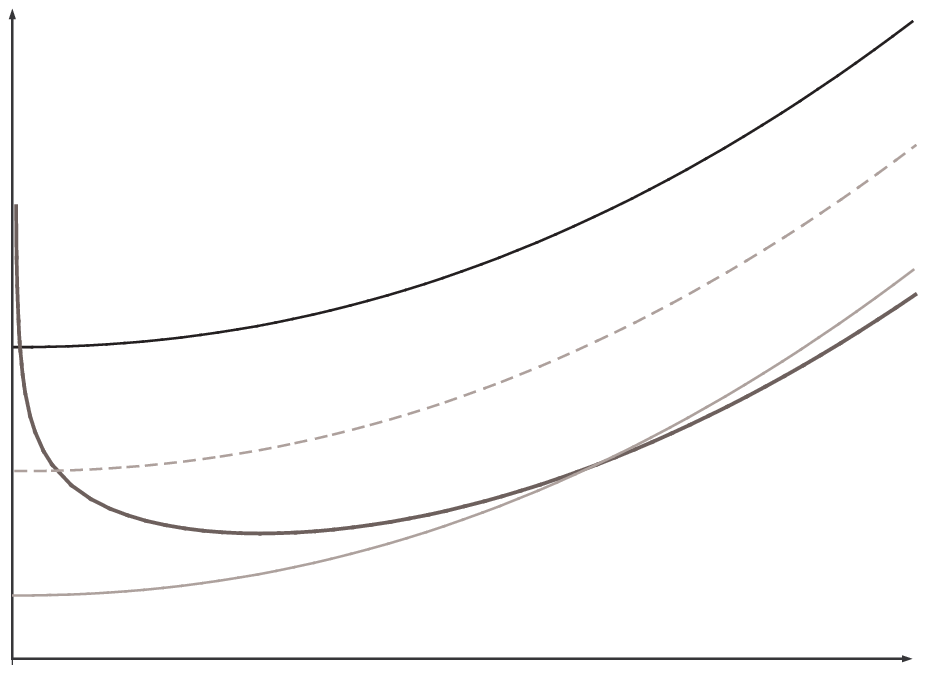}
\setlength{\unitlength}{1mm}
\put(-95,72){$v$} \put(-2,3.5){$x$}\put(-99,35.7){$v_1$}\put(-99,22){$v_2$}\put(-99,9){$v_3$}
\caption{\label{fig:inclinedplane}}{The intersection of two plots for $v_d$ and $v=\sqrt{v_0^2+2g\alpha x^k}\ (k>2)$ is the point of departure. Three curves for three different initial velocities $v_1$, $v_2$, and $v_3$ are plotted.}
\end{center}
\end{figure}
\subsubsection{$k=2$}
The case $k=2$ corresponds to a parabolic surface which we have considered earlier. If $v_0=0$, the particle remains unstable at the origin. For this case $A(x)=2\alpha v_0^2 $, and $B(x)=g$. If  $v_0> v_c:= \ds{\sqrt{\frac{g}{2\alpha}}}$ upon throwing the particle immediately leaves the parabolic surface.
If $v_0= v_c$ the particle moves tangent to the
parabolic surface and $N$ is always equal to zero. Finally if $v_0< v_c$, the particle moves on the parabolic surface and it never leaves the surface. 

\subsubsection{$1<k<2$}
For the case $1<k<2$, $\lim_{x\to 0} f''(x)\to -\infty$, and the radius of curvature at $x=0$, is equal to zero. So if the particle moves on the surface, near the origin $\ddot y\to -\infty$, which means that
$N$ should be $-\infty$, which is impossible. So for any arbitrary nonzero value of initial velocity, the particle leaves immediately the surface. This should be expected as this curve is beneath the parabola which is subject to the projectile motion. If $v_0=0$, the particle remains unstable at the origin.
\subsubsection{$k=1$}
For this case $A(x)=-\alpha^2 g$, and $B(x)=g$, never cross each other. In fact this case corresponds to an inclined plane. The particle which is thrown tangentially on the plane never leaves it.
\subsubsection{$k<1$}
$A(x)<0$ and $B(x)=g$ never cross each other. In this case $f''(x)>0$ for all $x$ and according to (\ref{07}) there is no point of departure. If  $v_0=0$, the particle starts moving but never leaves the surface.

\section{Particle  on a surface with friction}
In this section we extend the problem to include frictional
effects. The kinetic frictional force is assumed to be $f =\mu N$ opposing
the motion in the tangent plane of the surface, where $N$ is the normal force, and
the coefficients of kinetic  is $\mu$. The coefficient of static friction is also assumed to be $\mu$.

We put the particle at the point $x_0$ on the surface. It is easy to show that at any point $x_0$ on the surface where  $\mu> |f'(x_0)|$, the particle remains there at rest.
\subsection{Example; $f(x)=-\alpha x^k,\ (\alpha, k>0) $}
Let's find the point at which if we put the particle, it will remain at rest. For the case $k>1$, there is a point with $x$ component $a_1:=\left(\frac{\mu}{\alpha k} \right)^{1/(k-1)}$ on the surface. If the particle is put at any point $x\leq a_1$, it will remain at rest on that point.  For the case $k<1$, there is a point with $x$ component $a_2:= \left(\frac{\alpha k}{\mu} \right)^{1/(1-k)}$ on the surface. If the particle is put at any point $x\geq a_2$, it will remain at rest on the surface.
\subsubsection{$k>2$}
 As  we saw, it is not possible to throw the particle from the origin with any bounded initial velocity such that it leaves immediately the surface. If $v_0=0$, the particle remains at rest at the origin. There is a critical velocity, $v_{1c}$. If $v_0\leq v_{1c}$, the particle moves on the surface until it stops. At most it  reaches the point whose $x$ component is $a_1:=\left(\frac{\mu}{\alpha k} \right)^{1/(k-1)}$ and will be at rest there. When there is no friction the particles velocity is an increasing function and  eventually it leaves the surface. See figure 4. When there exists friction then the mechanical energy is not conserved, and one should take into account the work done by friction. Then the velocity of the particle  $v$ is less than  $\sqrt{v_0^2+2g\alpha x^k}$. So depending on $\mu$ and $v_0$ the two plots for $v_d(x)$ and $v(x)$ may have no intersection.
\subsubsection{$k=2$}
As it is shown if $v_0> v_c:= \ds{\sqrt{\frac{g}{2\alpha}}}$ upon throwing the particle,  it immediately leaves the parabolic surface. There is another critical velocity $v_{2c}$ which should also be less than $\ds{\sqrt{\frac{g}{2\alpha}}}$. If $v_0\leq v_{2c}$
then the particle moves on the surface until it reaches to a point $a'_1\leq \ds{\frac{\mu}{2\alpha }}$, and will be at rest there. If $v_{2c}<v_0< v_{c}$, then the particle passes $\ds{\frac{\mu}{2\alpha} }$, but it never leaves the surface.
\subsubsection{$1<k<2$}
For any arbitrary nonzero  initial velocity the particle leaves immediately the surface.
If the particle is put at any point $x\leq a_1:=\left(\frac{\mu}{\alpha k} \right)^{1/(k-1)}$ on the surface, it will remain at rest on that point.
\subsubsection{$k=1$}
This case corresponds to an inclined plane. The particle may remains at rest provided that $\mu>\alpha$. If the particle is thrown tangentially with any initial velocity, it will move on the plane for a finite time and eventually it will stop. If $\mu<\alpha$ the particle will slide on the plane forever.
\subsubsection{$k<1$}
If the particle is thrown tangentially with any initial velocity, it will move on the plane for a finite time and eventually it will stop.
 \begin{acknowledgments}
I would like to thank M. Khorrami, and A. H. Fatollahi for useful comments. This work is supported by the
research council of the Alzahra University.
\end{acknowledgments}

\end{document}